\documentclass[aps,prd,twocolumn,showpacs,amsmath,amssymb,nofootinbib]{revtex4-1}
\usepackage{graphicx}
\usepackage{color}
\usepackage{times}
\usepackage{inputenc}
\usepackage{bm}
\usepackage{ulem}
\usepackage{multirow}
\usepackage{float}
\usepackage{url}
\usepackage{natbib}
\usepackage{mathrsfs}
\usepackage{physics}
\usepackage{comment}
\usepackage[table,xcdraw]{xcolor}
\usepackage{booktabs}
\usepackage{comment}
\usepackage[caption=false]{subfig}
\usepackage[colorlinks=true,citecolor=blue,urlcolor=blue,linkcolor=blue]{hyperref}

\usepackage{tikz,xcolor,hyperref}
\definecolor{lime}{HTML}{A6CE39}
\DeclareRobustCommand{\orcidicon}{%
	\begin{tikzpicture}
	\draw[lime, fill=lime] (0,0) 
	circle [radius=0.16] 
	node[white] {{\fontfamily{qag}\selectfont \tiny ID}};
	\draw[white, fill=white] (-0.0625,0.095) 
	circle [radius=0.007];
	\end{tikzpicture}
	\hspace{-2mm}
}
\foreach \x in {A, ..., Z}{%
	\expandafter\xdef\csname orcid\x\endcsname{\noexpand\href{https://orcid.org/\csname orcidauthor\x\endcsname}{\noexpand\orcidicon}}
}

\begin{document}
\title{Exploring Radial Oscillations in Slow Stable and Hybrid Neutron Stars}

\author{Sayantan Ghosh$^{1}$\orcidA{}}
\author{Sailesh Ranjan Mohanty$^{1}$\orcidB{}}
\author{Tianqi Zhao$^{2}$\orcidC{}}
\author{Bharat Kumar$^{1}$\orcidD{}}
\email{kumarbh@nitrkl.ac.in}

\affiliation{\it $^{1}$ Department of Physics and Astronomy, National Institute of Technology, Rourkela 769008, India}
\affiliation{\it $^{2}$Department of Physics and Astronomy, Ohio University, Athens, Ohio 45701, USA}
\date{\today}
\begin{abstract}
In the era of gravitational wave astronomy, radial oscillations hold significant potential for not only uncovering the microphysics behind the internal structure but also investigating the stability of neutron stars (NSs). We start by constructing families of static NSs following nucleonic, quarkyonic, and hybrid equations of state and then subject them to radial perturbations in order to explore the stability of these stars. Unlike other literature where the fluid elements are assumed to be in chemical equilibrium, we consider the out-of-equilibrium effects on the chemical composition of fluid elements for the calculation of radial modes. Taking these considerations into account, we observe that the sound speed ($c^2_s$) and adiabatic index ($\gamma$) avoid singularities and discontinuities over the equilibrium case. We elucidate the response of the fundamental radial modes by examining the out-of-equilibrium matter distribution scenario, offering insights into its dynamic variations. We also demonstrate that this approach extends the stable branches of stellar models, enabling stars to sustain stable higher-order mass doublets, shedding some light on observation and existence of PSR J0740+6620.
\end{abstract}
\maketitle
\section{Introduction}
\label{intro}

Neutron stars (NSs), remnants of massive star cores, provide a unique laboratory for nuclear physics due to their dense composition. Pulsations called quasi-normal modes (QNMs)\cite{Kokkotas1999-vo, qnm2, qnm3, qnm4, qnm5} provide insights into the star's internal composition, with pressure as the primary restoring force. Studying QNMs helps probe the equation of state (EoS) of dense nuclear matter, crucial for understanding NS behaviour. While radial oscillations can not directly emit gravitational waves (GWs), their interaction with non-radial oscillations can amplify GW emissions, and create a stronger GW that might be detected \cite{passamonti2006coupling, PhysRevD.75.084038}. The emission of short gamma-ray bursts (SGRBs)\cite{sgrb1,sgrb2,sgrb3,sgrb4} in the process of forming hyper-massive NSs through binary NS mergers, can be influenced by the modulation of radial oscillations \cite{Chirenti_2019}.
Consequently, it holds significance not only for comprehending the physics of dense nuclear matter within a NS but also holds significance in the realm of gravitational wave (GW)\cite{doi:10.1126/science.aap9811,Abbott_2017,PhysRevX.9.011001,Abbott_2020} physics.

The Sturm–Liouville problem concerning the linear radial oscillations of variable stars under Newtonian gravity was thoroughly investigated by Ledoux and Walraven in 1958 \cite{wrubel1958variable}. When the square of the eigenfrequency in the radial oscillation equations becomes negative, the fluid displacement shows exponential growth, resulting in stellar instability. Chandrasekhar \cite{chandrasekhar1964dynamical} extended these principles to derive and solve similar radial oscillation equations in the context of general relativity. This instability is anticipated to occur prior to the Schwarzschild limit, particularly relevant for compact objects like white dwarfs (WDs) and NSs.

Conversely, Wheeler et al. proposed another stability criteria based on the mass-radius diagram, determined by solving the TOV equations\cite{harrison1965gravitation}. They argued that when the total mass decreases with increasing the central density, e.g. between stable branches of WDs and NSs, the star is no longer stable under self-gravity, leading to a collapse\cite{harrison1965gravitation}. Remarkably, for barotropic monotonic EoSs, these two methods of determining stellar instability criteria were found to be equivalent \cite{bardeen1966catalogue}.

Zero-temperature catalyzed NSs can be accurately described using a barotropic EoS, where thermal entropy is disregarded, and the chemical composition is determined by beta equilibrium. This assumption holds true when the relaxation time for beta equilibrium is significantly shorter than the oscillation period, which ensures that NS matter remains catalyzed throughout compression and expansion oscillations. At lower temperatures, when the temperature is less than 1 MeV, the neutrino relaxation time becomes longer compared to the oscillation time, preventing the beta-equilibrium condition from being met during oscillations \cite{alford2019damping}. Consequently, matter composition becomes frozen under compression, leading to a higher adiabatic index. This effect substantially extends the range of maximally stable configurations for WDs, from $\rho_c = 1.2 \times 10^9$ g.cm$^{-3}$ to $\rho_c = 4 \times 10^{10}$ g.cm$^{-3}$, corresponding to the maximum mass configuration of a WD \cite{chanmugam1977radial}.

The influence of the slow weak process differs significantly between WDs and NSs due to variations in the beta-equilibrium composition, which ranges from a large proton fraction of approximately 0.5 on the stellar surface to approximately 0.05 around the crust-core transition density. The assumption of a small proton fraction at the center of a NS aligns the adiabatic index of the beta-equilibrium matter with that of matter involved in a slow weak process. However, this assumption may not hold in EoS models where the proton fraction rises in the core of a NS, reaching values greater than 0.2 \cite{zhao2020quarkyonic}.

Furthermore, when a phase transition occurs, the adiabatic index differs significantly between catalyzed matter and matter with a fixed chemical composition \cite{jaikumar2021g}. Specifically, when there is a density discontinuity due to a first-order phase transition between hadronic matter and quark matter, this effect can be manifested through slow and rapid junction conditions \cite{pereira2018phase}, as well as intermediate junction conditions \cite{pereira2018phase,rau2023two,rau2023nonequilibrium}. Although the barotropic EoS is employed for both hadronic matter and quark matter, it results in two distinct compositions with an infinite composition gradient \cite{constantinou2023framework}. This significant composition gradient supports non-radial g-mode oscillations, which may be observable in the third generation of gravitational observations\cite{zhao2022quasinormal}. In the case of radial oscillations, a large density discontinuity with slow junction conditions leads to a long branch of a slow-stable hybrid star \cite{lugones2023model,ranea2023asteroseismology}, while sequential density discontinuities result in multiple slow-stable branches \cite{gonccalves2022impact,rau2023two,rau2023nonequilibrium}.

This work is the first study of slow, stable NSs (hadronic and hybrid) with no density discontinuity, where radial oscillations are solved under the assumption of a fixed chemical composition in the slow transition limit. However, previous works on non-equilibrium (adiabatic) radial oscillations focus on hadronic NSs \cite{chanmugam1977radial, gourgoulhon1995maximum} or hybrid NSs with a first-order phase transition \cite{lugones2023model, ranea2023asteroseismology, rau2023two, rau2023nonequilibrium}.

This paper is organized as;
in Section (\ref{sec:form}), we describe theoretical formalism, where in Sub-sections \ref{ss}, \ref{subsec:TOV}, and \ref{subsec:radosc}, we have discussed sound speed, hydrostatic equilibrium, and radial oscillation, respectively. We present our results and discussion part in Section (\ref{results}), and finally in Section (\ref{con}) we conclude our work.

Throughout this paper, we adopt mostly positive signatures (-, +, +, +) and utilize a geometrized unit system (G=c=1).

\section{Theoretical Framework}
\label{sec:form}

We utilize the Zhao-Lattimer (ZL) model \cite{zhao2020quarkyonic} to characterize nucleonic matter. The nuclear saturation properties is set as: saturation density ($n_{sat}$) at 0.16 fm$^{-3}$, the binding energy ($E_{sat}$) at -16 MeV, the compression modulus ($K_{sat}$) at 230 MeV, and the symmetry energy ($S_v$) at 31 MeV and the symmetry energy slope ($L$) at 70 MeV, while keeping the power-law index ($\gamma_1$) constant at a value of 2.

For the quark EoS, we use the vMIT bag model~\cite{Gomes:2018eiv, Klahn:2015mfa} with Lagrangian density
\begin{eqnarray}
\mathcal{L}&=&\sum_{q=u,d,s}\left[\bar{\psi}_{q}\left(i \gamma^\mu\partial_\mu -m_{q}-B\right) \psi_{q}\right]+\mathcal{L}_{\mathrm{vec}} ,\\
\mathcal{L}_{\text {vec }}&=&-G_{v} \sum_{q} \bar{\psi} \gamma_{\mu} V^{\mu} \psi+\left(m_{V}^{2} / 2\right) V_{\mu} V^{\mu} .
\end{eqnarray}
where $V_{\mu}$ represent vector meson field with repulsive interactions $(G_v/m_V)^2= 0.2$ fm$^{-2}$. $B^{1/4}=180$ MeV is a constant reflecting the cost of confining the quarks inside the bag, and the $m_{q}$ are the current quark masses (here, $m_u = 5$ MeV, $m_d = 7$ MeV, and $m_s = 150$ MeV).

Radial oscillation and stability of slow stable hybrid stars have been studied under first-order phase transition with Maxwell construction. Here we study hybrid NS with crossover phase transition\cite{kapusta2021neutron} and first order phase transition with Gibbs construction\cite{constantinou2023framework}. Details for the construction of hybrid EoS are described in \cite{PhysRevD.105.103025}.

\subsection{Sound speed}
\label{ss}

Phase transitions or crossovers occurring in dense matter within NSs have a direct impact on the speed of sound, $c_s$($n_B$), in relation to baryon density ($n_B$). The alterations in internal composition or structure, prompted by extreme conditions, result in unique signatures discernible in the propagation of sound \cite{PhysRevD.107.014011,Ecker_2022,chatterjee2023analyzing,PhysRevC.101.045803,Altiparmak_2022,Bedaque_2015,Ferrer_2023,yao2023structure}. The examination of variations in the speed of sound serves as a crucial tool for probing EoSs and gaining a comprehensive understanding of the fundamental properties of NSs, including the exotic states of matter they contain.

The adiabatic sound speed is obtained by
\cite{PhysRevD.105.103025} 
\begin{equation}
    c^2_{ad}(n_B,y_i) = \left(\frac{\partial{P}}{\partial{n_B}}\right)_{y_i}\left(\frac{\partial{\mathcal{E}}}{\partial{n_B}}\right)_{y_i}^{-1}
\end{equation}
where $P$ and $\mathcal{E}$ are the total pressure and energy density respectively and 
$y_i$ is the particle fraction in favour of the total baryon number density($n_B$) [$y_i\rightarrow y_{i,\beta}(n_B)$, $i = n,p,u,d,s,e,\mu$].
Then evaluating it in $\beta$-equilibrium
\begin{equation}
    c^2_{ad,\beta}(n_B) = c^2_{ad}[n_B,y_{i,\beta}(n_B)].
\end{equation}
The equilibrium sound speed is expressed as the total derivative of the pressure and energy density concerning the baryon density, following the implementation of $\beta$-equilibrium,
\begin{equation}
    c^2_{eq} = \left(\frac{dP_\beta}{dn_B}\right)\left(\frac{d\mathcal{E}_\beta}{dn_B}\right)^{-1}
\end{equation}


\subsection{Hydrostatic equilibrium }
\label{subsec:TOV}

A non-rotating NS's intense gravitational forces lead to a nearly perfect spherical equilibrium. This symmetry allows for a reasonable assumption of spherical symmetry. The Schwarzschild metric accurately describes the gravitational field of such a body\cite{1916AbhKP1916..189S}:
\begin{equation}
ds^2 = -e^{2\nu}  dt^2 + e^{2\lambda} dr^2 + r^2(d\theta^2 + \text{sin}^2\theta d\phi^2),
\label{eqn4}
\end{equation}
where $\lambda \equiv \lambda (r)$ and $\nu \equiv \nu (r)$ are metric functions, each following their respective set of equations. Here, the energy-momentum tensor $T_{\mu\nu}$ takes the form of a perfect fluid:
\begin{equation}
T_{\mu\nu}=(P+\mathcal{E})u_\mu u_\nu + Pg_{\mu\nu},
\label{eqn14}
\end{equation}
where 
$P$ is pressure, 
$\mathcal{E}$ is energy density, and 
$u_\mu$ represents covariant velocity. In the context of spherical symmetry, the components $u_0$ and $u_1$ exhibit non-zero values along the radial direction.

Applying Einstein's field equations to the Schwarzschild metric in eq. (\ref{eqn4}) under equilibrium conditions and utilizing the boundary condition $\lambda(r=0)=0$, we obtain:
\begin{equation}
e^{-2\lambda(r)} = \left(1-\frac{2m}{r}\right),
\label{eqn7}
\end{equation}
The mass $m$ can be determined through integration using:
\begin{equation}
\frac{dm}{dr} = 4\pi r^2 \mathcal{E} .
\label{eqn6}
\end{equation}
Likewise, employing the law of conservation of momentum, we obtain \cite{landau1967classical}:
\begin{equation}
\frac{d\nu}{dr}=-\frac{1}{P+\mathcal{E}}\frac{dP}{dr}.
\label{eqn8}
\end{equation}
Ultimately, employing equation (\ref{eqn8}) and Einstein's field equations leads to \cite{oppenheimer1939massive, tolman1939static}:
\begin{equation}
\frac{dP}{dr} = - \frac{m}{r^2} \frac{\left( P+\mathcal{E} \right) \left( 1 + \frac{4\pi r^3P}{m}\right)}{\left(1-\frac{2m}{r}\right)}.
\label{eqn5}
\end{equation}
Eqs. (\ref{eqn7}) and (\ref{eqn8}) describe the behaviour of the metric functions within the NS where $r<R$. At the surface, i.e. at $r=R$, they satisfy the boundary condition, 
\begin{equation}
e^{2\nu(R)} = e^{-2\lambda(R)} = \left(1-\frac{2M}{R}\right).
\label{eqn9}
\end{equation}
Eq. (\ref{eqn9}) remains valid even beyond the star, where $R$ is replaced by $r$ for $r > R$, taking on the familiar form of the Schwarzschild solution.


\subsection{Radial oscillation equations}
\label{subsec:radosc}

Maintaining the spherical symmetry of the background equilibrium configuration, we perturb both fluid and spacetime variables. We assume a harmonic time dependence for the radial displacement of the fluid element at position $r$ in the unperturbed model,
\begin{equation}
    \delta r (r,t) = X(r)e^{i\omega t},
\end{equation}
The equations for linearized radial perturbations can be expressed as \cite{chandrasekhar1964dynamical,kokkotas2001radial}
\begin{widetext}
\begin{equation}\label{RPE}
\begin{aligned}
c_{s}^{2} X^{\prime \prime}+\left(\left(c_{s}^{2}\right)^{\prime}-Z+ {4 \pi} r \gamma P e^{2 \lambda}
-\nu^{\prime} \right) X^{\prime}& \\
+ {\left[2\left(\nu^{\prime}\right)^{2}+\frac{2 m}{r^{3}} e^{2 \lambda}-Z^{\prime}-4 \pi(P+\mathcal{E}) Z r e^{2 \lambda} + \omega^{2} e^{2 \lambda-2 \nu}\right]X} &= 0,
\end{aligned}
\end{equation}
\end{widetext}
Here, primes represent differentiation with respect to the radial coordinate $r$, and $c_s^{2}$ denotes the square of the speed of sound. The adiabatic index, $\gamma$, is related to the speed of sound as follows:
\begin{equation}
\gamma=\left(\frac{P+\mathcal{E}}{P}\right)c_s^{2} .
\end{equation}
Now we can define the $\gamma$ by considering the two different kinds of sound speeds, as
\begin{equation}
\gamma_{eq}=\left(\frac{P+\mathcal{E}}{P}\right)c_{eq}^{2}, \\
\gamma_{ad}=\left(\frac{P+\mathcal{E}}{P}\right)c_{ad}^{2},
\end{equation}
where the suffix "eq" and "ad" correspond to the equilibrium and adiabatic case respectively.
\begin{equation}
Z(r)=c_{s}^{2}\left(\nu^{\prime}-\frac{2}{r}\right).
\end{equation}
Now, we re-define the displacement function as 
\begin{equation}\label{zeta}
    \zeta=r^{2}e^{-\nu}X.
\end{equation}
Eq. (\ref{RPE}) can be rewritten for $\zeta$ as 
\begin{equation}\label{RPEs}
\frac{d}{dr}\left( H\frac{d \zeta}{d r}\right)+\left(\omega^{2} W+Q\right) \zeta=0,
\end{equation} 
with
\begin{subequations}\label{rf}
\begin{align}
\label{rf1}
H&=r^{-2}(P+\mathcal{E}) e^{\lambda+3 \nu} c_{s}^{2} \\
\label{rf2}
W&=r^{-2}(P+\mathcal{E}) e^{3 \lambda+\nu} \\
\label{rf3}
Q&=r^{-2}(P+\mathcal{E}) e^{\lambda+3 \nu}\left(\left( \nu^\prime \right)^{2}+\frac{4}{r} \nu^{\prime}- {8 \pi} e^{2 \lambda} P\right) .
\end{align}
\end{subequations}

$H$, $W$, and $Q$ are dependent on the radial coordinate $r$ and can be computed using the unperturbed background configuration. Notably, eq. \eqref{RPEs} explicitly reveals its self-adjoint nature. The Lagrangian variation of the pressure now is
\begin{equation}\label{dp}
    \Delta P = -r^{-2} e^{\nu}(P+\mathcal{E}) c_{s}^{2}\zeta^{\prime}.
\end{equation}
As radial oscillations cannot displace the fluid element at the center, the boundary condition at the center is
\begin{equation}\label{BC1}
    X(r=0)=0.
\end{equation}
 At the stellar surface, the Lagrangian variation of pressure should vanish, expressed as
\begin{equation}\label{BC2}
\Delta P(R) = 0.
\end{equation}

The differential equation \eqref{RPEs}, subject to the boundary conditions in eqs. \eqref{BC1} and \eqref{BC2}, constitutes a Sturm-Liouville eigenvalue problem \cite{galaxies11020060,Hlad_k_2020,Pinku_prd_2023,Pinku_mnras_2023}. 
\begin{figure*}
    \includegraphics[width=0.49\textwidth, height= 0.43\textwidth]{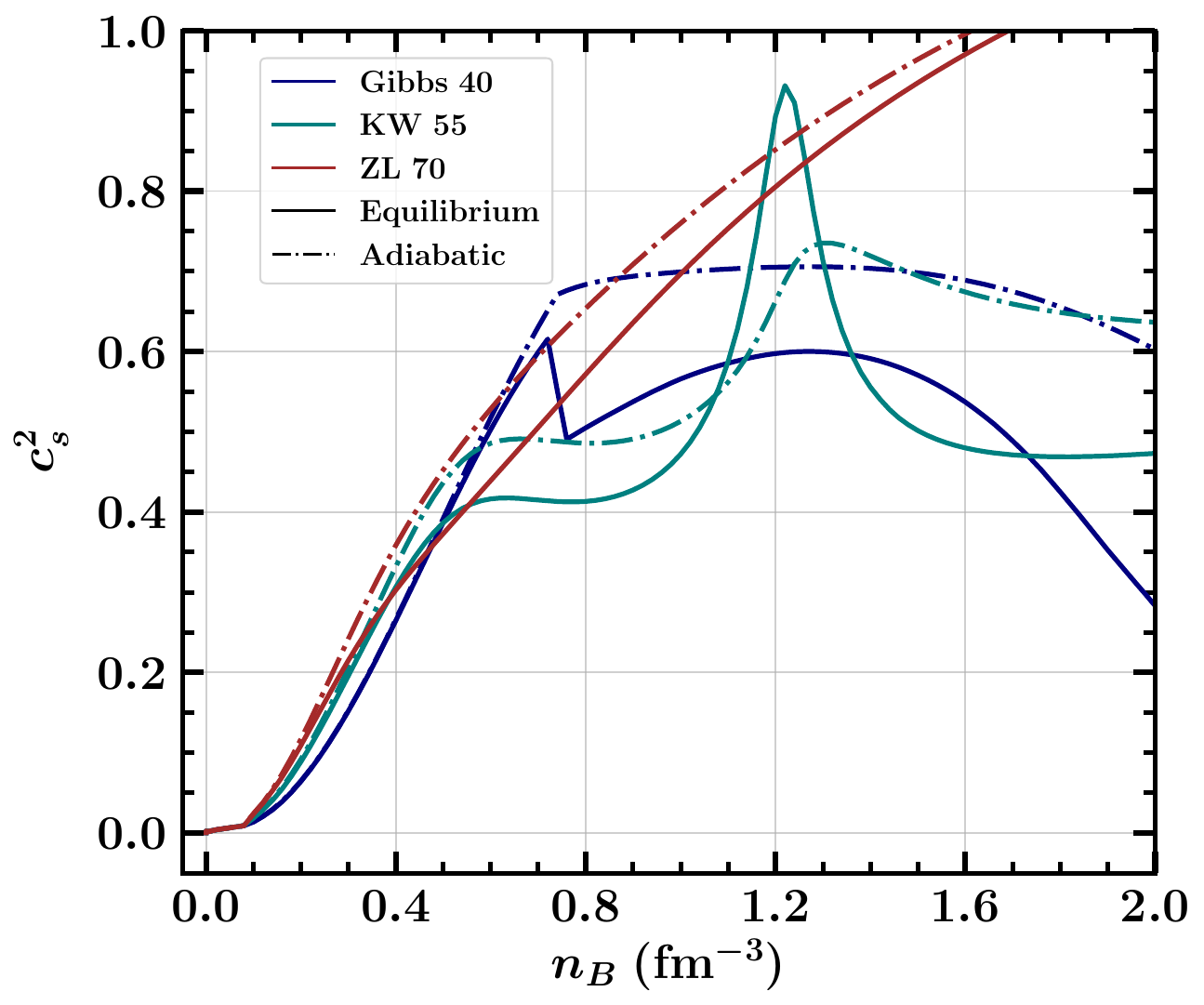}
    \includegraphics[width=0.49\textwidth, height= 0.43\textwidth]{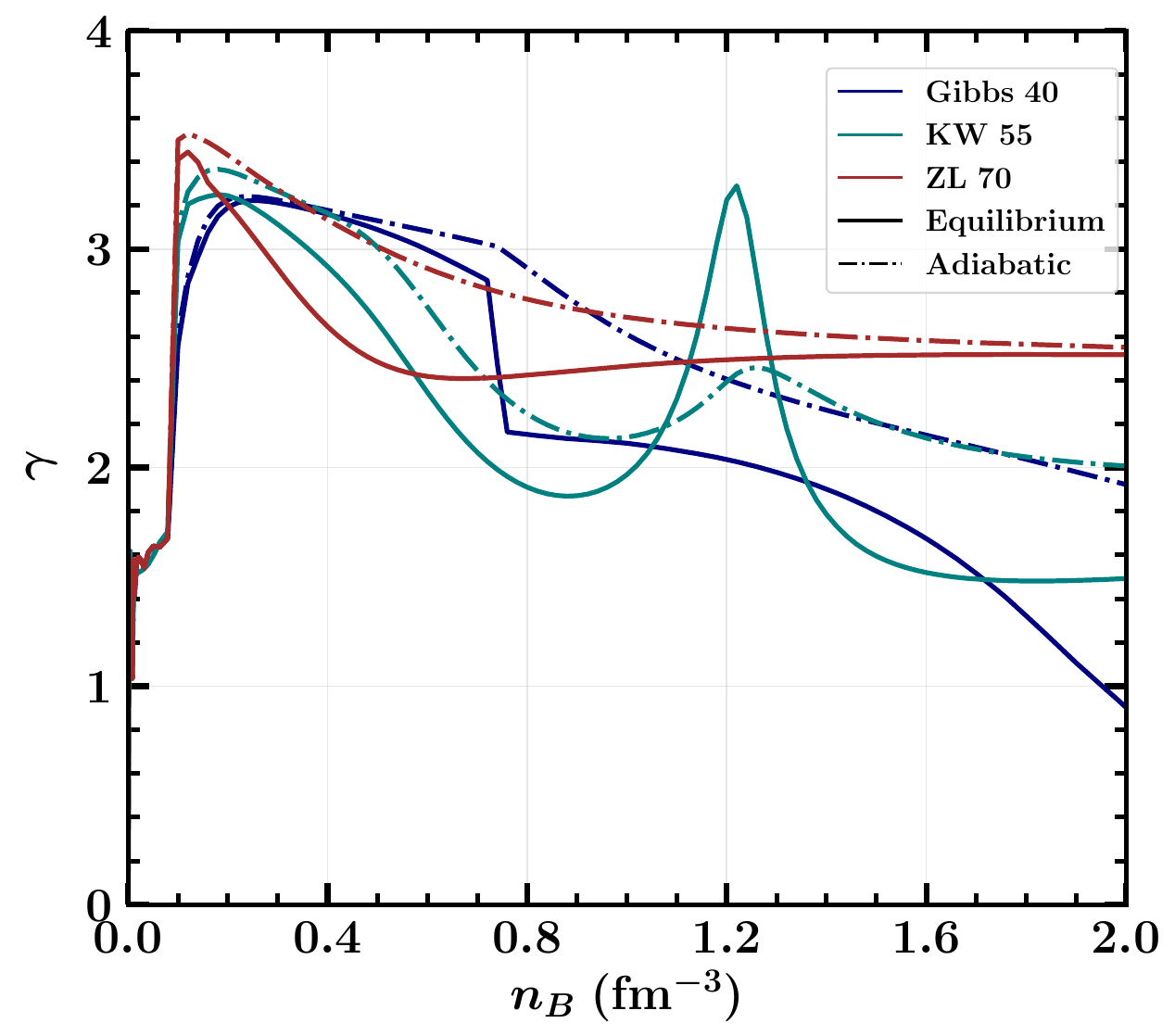}
    \caption{
On the left panel, the squared sound speeds ($c^2_s$) are presented as functions of baryon number density ($n_B$), delineating the relationship between the local speed of sound and the density of baryonic matter. The right panel exhibits the adiabatic index ($\gamma$) as a function of baryon number density ($n_B$) for the specified models detailed in the inset. Solid lines represent the equilibrium case, while dashed-dot lines depict the adiabatic scenario.}
    \label{soundspeed}
\end{figure*}

\begin{figure*}
    \centering
    \includegraphics[width=0.49\linewidth]{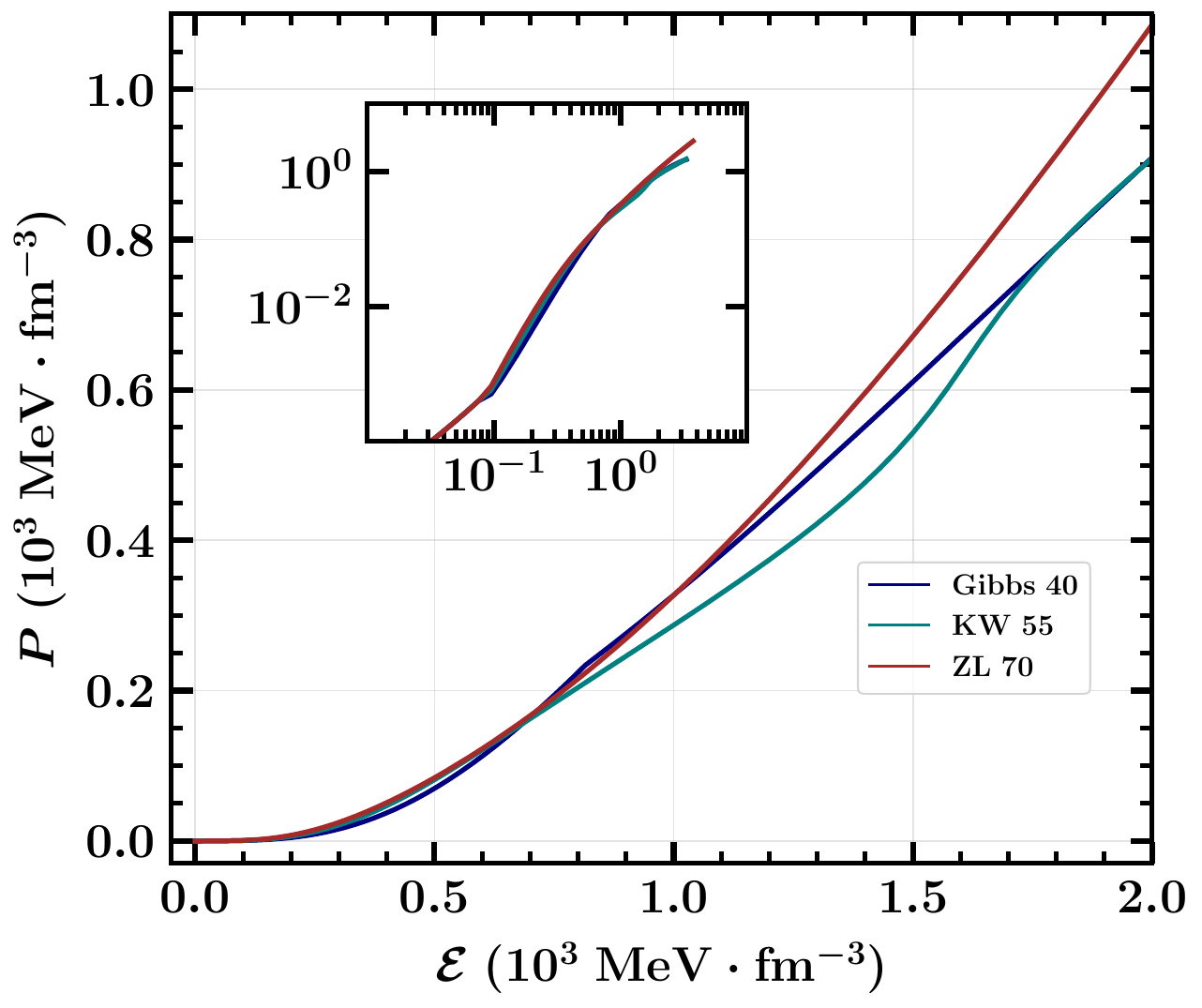}
    \includegraphics[width=0.49\linewidth]{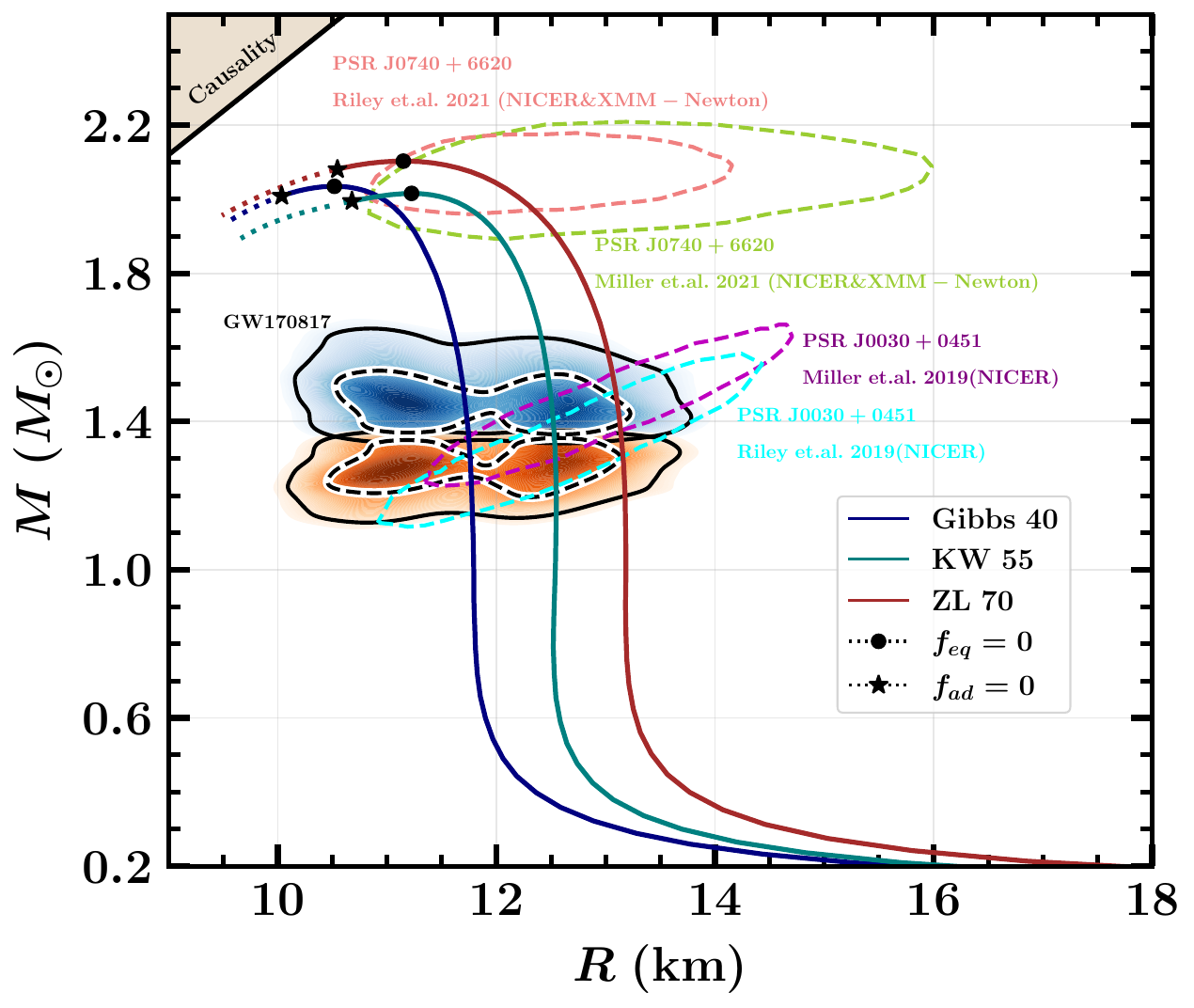}
    \caption{On the left panel EoSs are represented as pressure $P$ versus energy density $\mathcal{E}$. The composition of the three employed models is as follows: for ZL, nucleons and leptons; for Gibbs, nucleons, quarks, and leptons in a soft first-order phase transition description; and for KW, the same as for Gibbs, but in a cross-over description. The right panel displays the Mass-Radius relationship obtained for our choice of EoSs. The astrophysical observable constraints on mass and radius from PSR J0740+6620 \cite{Miller_2021, Riley_2021}, and NICER data for PSR J0030+0451 \cite{Miller_2019, Riley_2019} are represented by coloured regions. The outer and inner regions of the blue and orange butterfly structured plot indicate the 90\% (solid) and 50\% (dashed) confidence intervals based on the LIGO-Virgo analysis for Binary Neutron Star (BNS) components of the GW170817 event\cite{GW170817, PhysRevLett.120.172702, PhysRevC.107.055804}.}
    \label{eos_mr}
\end{figure*}
\begin{figure*}[t!]
    \includegraphics[width=\linewidth]{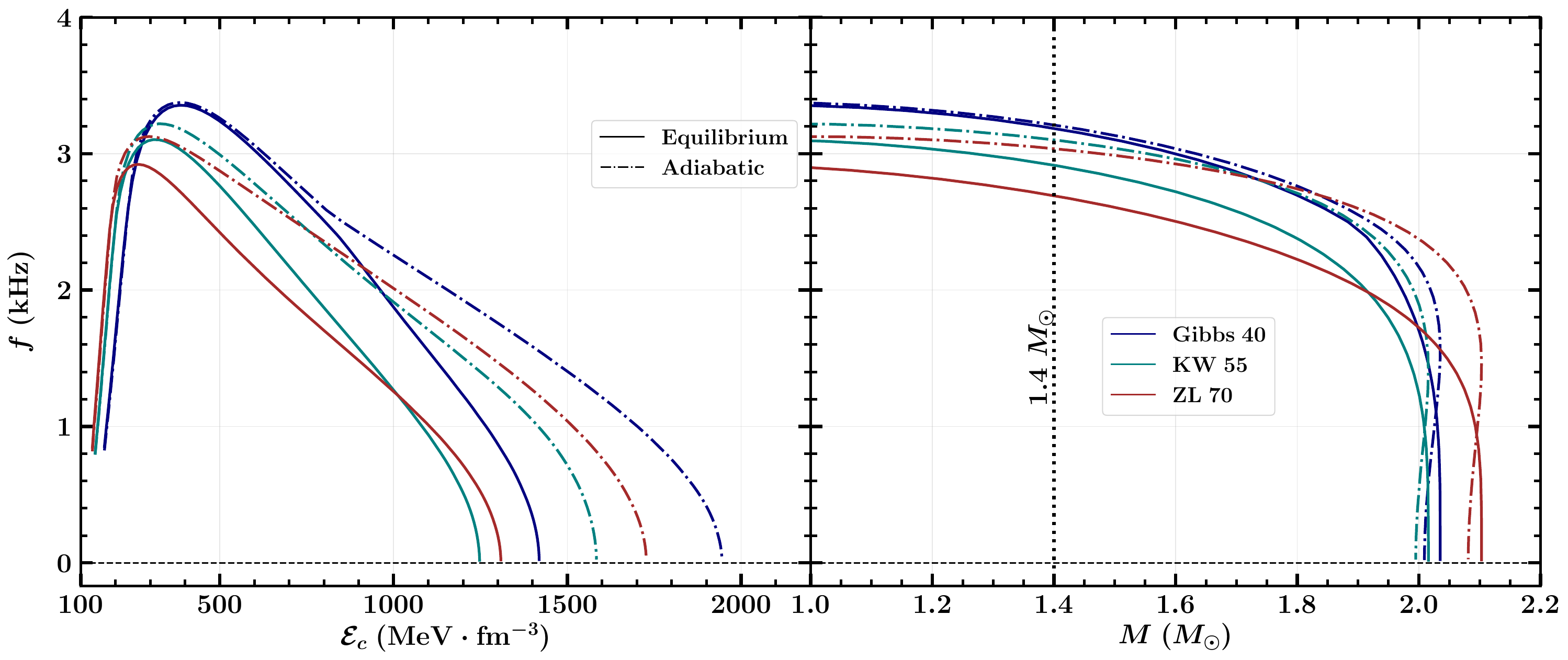}
    \caption{
    On the left panel, the  $f$-mode frequency is depicted as a function of central energy density ($\mathcal{E}_c$), offering insights into the dynamic behaviour in response to varying internal energy conditions. The right panel illustrates the   $f$-mode frequency against the mass ($M$). 
    }
    \label{fmode}
\end{figure*}
\section{Results and Discussions}
\label{results}
In this section, we discuss numerical results obtained by considering three different EoSs such as Gibbs40, KW55, and ZL70.
Firstly, we have studied the variation of the sound speed and adiabatic index under different conditions. Additionally, our investigation focuses on the mass-radius relation of NSs. Furthermore, our attention extends to the $f$-mode oscillation, to understand the dynamics of NS.

In the left panel of Fig. \ref{soundspeed}, we depict the variation of sound speed with baryon density. Notably, the curves corresponding to the purely nucleonic model (ZL) consistently exhibit a steep rise as baryon density increases. Conversely, when quarks are introduced, as illustrated by the KW and Gibbs models, a distinct behaviour emerges, showing up less dramatically,
but still visibly in terms of a more or less pronounced
change of slope in $c_s$($n_B$). The presence or enhancement of quarks introduces a notable deviation from the monotonic trend observed in the purely nucleonic scenario.

In the case of the KW model, the growth in sound speed is either slowed down or reversed, creating a smoother transition compared to the Gibbs model. For the Gibbs model, the effect is more abrupt and discontinuous. The enhancement of quarks in the Gibbs model leads to a sharp alteration in the growth pattern of sound speed, introducing a distinctive discontinuity in the curves.

When calculating the equilibrium sound speed, we are doing the total pressure derivative concerning the energy density. However in the adiabatic case, it is different, here we are calculating the corresponding sound speed for each composition. So
one noteworthy observation is, that the curves become slightly smoother when considering the adiabatic case \cite{PhysRevD.104.123032}, in comparison to the equilibrium case. Now, examining the right panel of Fig. \ref{soundspeed}, we observe a similar effect on the adiabatic index ($\gamma$) as seen in the left panel, reflecting the influence of quark content on both sound speed and $\gamma$.

The Mass-Radius relation, depicted in the right panel of Fig. \ref{eos_mr}, illustrates the diverse behaviour of NSs under different EoSs. At lower masses the NS radius is larger, reflecting the pressure support provided by the nuclear and sub-nuclear constituents resisting gravitational collapse. As the mass increases, the star's gravity becomes stronger, leading to a more compact configuration with a smaller radius.
However, there is a limit to how much mass a NS can support against gravity. This upper limit is the maximum mass, shown in the M-R curve with dot markers. If the mass exceeds this limit, the NS is unstable in the equilibrium case but is still stable in the adiabatic case. Star markers denote the NS becomes unstable even in adiabatic cases. In the regions between the star and the dot markers are known as the slow stable NS. In regions beyond the star markers (the dotted portion), the NS go through gravitational collapse into a Black hole (BH).
In this representation, distinct markers convey specific characteristics:
\begin{itemize}
    \item The star marker signifies the point where the $f$-mode frequency (adiabatic) tends to zero, occurring beyond the maximum mass configuration.
    \item The dot marker corresponds to the $f$-mode frequency (equilibrium) reaching zero, typically aligning with the configuration of maximum mass.
\end{itemize}
So to study the instability in NS we have to consider the adiabatic case since it is in the unstable region.
Upon closer examination of the figure, a notable observation is that the maximum mass associated with the ZL EoS surpasses that of the other two EoSs. The Gibbs EoS exhibits a marginally higher maximum mass compared to the KW EoS. These discernible differences in maximum mass values are quantified in Table \ref{table:MR_s}.
The differences between the curves are due to the absence (ZL is purely hadronic) or presence (Gibbs, KW) of quarks, and quark matter can be more compressible than nuclear matter, leading to a softer EoS.
\begin{table}
    \centering
    \caption{NS parameters of the maximally stable NS corresponding to both equilibrium and adiabatic cases for chosen EoSs. Central energy density $(\mathcal{E}_c)$ is in $(\mathrm{MeV \cdot fm^{-3}})$, Mass in $(M_\odot)$ and Radius in $(\mathrm{km})$. \\}
    \setlength{\tabcolsep}{2mm}
    \renewcommand{\arraystretch}{1.5}
    \scalebox{0.85}{
        \begin{tabular}{cccccccc}
            \hline \hline
            & \multicolumn{3}{c}{$f_\mathrm{eq}=0$} && \multicolumn{3}{c}{$f_\mathrm{ad}=0$} \\
            \cline {2-4} \cline {6-8} 
            EoS & $\mathcal{E}_c$ & $M$ & $R$ && $\mathcal{E}_c$ & $M$ & $R$  \\
            \hline \vspace{-3mm} \\
            Gibbs40 & 1419 & 2.035 & 10.492 && 1944 & 2.009 & 10.035  \\
            KW55 & 1247 & 2.016 & 11.228 && 1583 & 1.995 & 10.678  \\
            ZL70 & 1308 & 2.103 & 11.180 && 1727 & 2.081 & 10.545 \\
            \hline \hline
        \end{tabular}}
    \label{table:MR_s}
\end{table}

In Fig. \ref{fmode}, the variation of the $f$-mode frequency with central energy density ($\mathcal{E}_c$) is presented in the Left panel, while the Right panel displays the variation with Mass ($M$), aiming to explore radial oscillation and stability. The curves for all EoSs exhibit a consistent trend. Notably, a closer examination of the Left panel reveals that at a specific central density (approximately 300 $\mathrm{MeV \cdot fm^{-3}}$) for Gibbs and KW, the $f$-mode frequency surpasses that of ZL, indicating the point at which quarks start to leak out from the nucleons. Now, let's examine the distinctive behaviour of the $f$-mode frequency curves corresponding to the equilibrium and adiabatic sound speed case, arising from the contrasting timescales for reaching equilibrium. considering the equilibrium sound speed case, represented by the "solid" line, the NS matter quickly approaches the ground state, rapidly converging to a new equilibrium state and causing the $f$-mode frequency to vanish at a lower central density. This immediate response signifies the prompt stabilization of the NS in the equilibrium scenario. Conversely, for the adiabatic sound speed case, depicted by the "dashed-dot" line, the process unfolds more gradually, necessitating a more extended duration for the NS to attain equilibrium. Consequently, the $f$-mode frequency diminishes at a higher central density, highlighting the prolonged timescale associated with the adiabatic case. Moving to the Right panel of this figure, in proximity to 1.4 $M_\odot$, the $f$-mode frequency of Gibbs surpasses that of KW, and for KW, it exceeds that of ZL [refer to Table \ref{table:MR_s}]. Remarkably, the $f$-mode frequency for ZL diminishes at a later stage compared to Gibbs and KW. This suggests that the hadronic EoS maintains constancy until it attains its maximum mass.
\section{Summary \& Conclusions}
\label{con}
In this study, we thoroughly examined three distinct EoSs, each distinguished by a unique symmetry energy slope (L): ZL70 representing a nucleonic EoS, and Gibbs40 \& KW55 representing hybrid EoSs, with KW55 incorporating crossover matter. The differences between primary focus centred around an in-depth analysis of the adiabatic sound speed in comparison to the equilibrium scenario. A prominent observation emerged, showing smoother trends in the sound speed dynamics when considering the adiabatic case over the equilibrium one. Similarly, the adiabatic index ($\gamma$) manifests smoother trends in the adiabatic case than in equilibrium.
 Then we studied the mass-radius relation of NS with those aforementioned EoSs and found some different behaviours like, ZL70 showcasing heightened stiffness compared to the other two. Additionally, Gibbs40 exhibited more stiffness than KW55. Within the Mass-Radius curve, we meticulously identified stable and unstable branches of different EoSs. Shifting our focus to radial oscillations, we probed the behaviour of the fundamental mode frequency ($f$-mode). Considering the different EoSs, Gibbs40 has a larger frequency than the other two at a particular mass (say 1.4 $M_\odot$) of NS. We have seen that for the equilibrium case, $f$- mode frequencies vanish at lower central energy densities, however, for the adiabatic case, it vanishes at much larger central energy densities. As a result, the stable branch can be extended beyond the maximum mass configuration in an adiabatic case. Although this slow stable branch has a similar mass as the maximum mass, the radius can be 0.6 km smaller. Since maximum mass configuration usually corresponds to the lower bound of radius, the slow stable branch extends the lower bound of NS radius, which is particularly relevant to radius measurement of high mass NSs such as PSR J0740+6620 \cite{Miller_2021, Riley_2021}. A radius measurement of $R<11$km for two solar mass neutrons might indicate a NS on the slow stable branch, while PSR J0740+6620 is more likely to be on the regular stable branch based on the NICER observation.
Our investigation sought to unravel how the $f$-mode frequency responds by considering the adiabatic sound speed case, providing insights into its dynamic changes. Our stability analysis, focused on linearized oscillation where the perturbation is small, acknowledges the potential influence of higher-order terms that could render slow stars unstable at large oscillation displacement, as discussed by Gourgoulhon et al. \cite{gourgoulhon1995maximum}.

\label{conclusion}

\section{Acknowledgments}
B.K. acknowledges partial support from the Department of Science and Technology, Government of India, with grant no. CRG/2021/000101. TZ is supported by the Network for Neutrinos, Nuclear Astrophysics and Symmetries (N3AS), through the National Science Foundation Physics Frontier Center Grant No. PHY-2020275.

\bibliographystyle{apsrev4-1}
\bibliography{RO.bib}
\end{document}